\documentclass{PoS}
\usepackage{amsmath}

\title{Volume Effects on the Method of Extracting Form Factors at Zero Momentum}

\ShortTitle{Form Factors $@$ Zero Momentum}

\author{\speaker{Brian C.~Tiburzi}
\thanks{Work supported in part by a joint City College of New York--RIKEN/BNL Research Center fellowship, 
an award from the Professional Staff Congress of the CUNY, 
and by the U.S.~National Science Foundation, 
under Grant No.~PHY-$1205778$. 
}\\
        Department of Physics,
        The City College of New York,  
        New York, NY, USA\\
        Graduate School and University Center,
        The City University of New York,
        New York, NY, USA\\
        RIKEN BNL Research Center, 
        Brookhaven National Laboratory, 
        Upton, NY, USA\\
        E-mail: \email{btiburzi@ccny.cuny.edu}}

\abstract{The Rome method allows one to extract form factors using lattice computations performed strictly at zero momentum.  We investigate the size of finite volume effects resulting from this method. As a test case, we focus on the pion charge radius and show how to ascertain the finite volume effect with the aid of chiral perturbation theory. The framework developed can easily be generalized to account for modified infrared physics of other low-energy matrix elements extracted at zero momentum.}

\FullConference{The 32nd International Symposium on Lattice Field Theory,\\
		23-28 June, 2014\\
		Columbia University New York, NY}

\begin{document}

\section{Introduction and Motivation}

Computation of momentum-dependent matrix elements between hadron states is required for many phenomenological applications. 
The nucleon electromagnetic form factors are such an example;
and, 
in particular, 
the magnetic moment, charge radius and magnetic radius cannot be accessed in standard lattice three-point function  calculations without the insertion of momentum transfer. 
Consequently these observables require a large extrapolation to zero momentum, 
and this forces one to rely on models for which the associated uncertainty becomes difficult to quantify. 
In the experimental extraction of radii, 
form factor data are similarly modeled, 
however, 
experiments generally have access to smaller momentum transfer than available to lattice QCD, 
which, 
for periodic boundary conditions, 
amounts to
$\frac{2 \pi}{L} \sim 300$--$500 \, \texttt{MeV}$. 
While twisted boundary conditions can overcome the restriction to quantized momenta%
~\cite{Bedaque:2004kc,deDivitiis:2004kq,Sachrajda:2004mi}, 
the partially twisted scenario is currently the only practicable option. 
Without the generation of new gauge configurations, 
however, 
there are strong correlations between results at differing twist angles. 
When such correlation are accounted for, 
statistically independent information generally cannot be garnered about the low-momentum behavior of hadronic matrix elements.

A novel method, 
which we refer to as the
\emph{Rome} 
method, 
has recently been proposed for the 
computation of form factors directly at zero momentum%
~\cite{deDivitiis:2012vs}. 
In essence, 
the method boils down to the computation of modified correlation functions which determine the Taylor series coefficients in an expansion about vanishing momentum. 
Computing the Taylor series coefficients directly removes the uncertainty associated with extrapolation to zero momentum, 
and the Rome method was demonstrated for two phenomenologically interesting applications:
form factors of flavor-changing currents at the end point, 
and the hadronic vacuum polarization at zero momentum. 
In looking to additional applications of the Rome method, 
we note that the Taylor-coefficient correlation functions 
require the evaluation of certain integrals over 
$n$-point functions, 
and so can be computationally quite expensive.%
\footnote{
An alternate approach to the Rome method is conceivable. 
Instead of computing the Taylor series coefficients directly, 
it is possible to approximate them by determining the quark propagator as a function of twist angle, 
and then numerically differentiating with respect to the twist. 
This alternative corresponds to studying the variation of the quark propagator with respect to a uniform gauge potential;
and, 
as such, 
is similar to studying the variation of hadronic matrix elements with respect to terms in the quark action
\`a la Feynman--Hellmann. 
An application of the latter has been proposed for the study of the spin-structure of hadrons%
~\cite{Chambers:2014qaa}.
} 
In advance of further numerical studies, 
it is valuable to ascertain the size of finite volume corrections to the Rome method. 
Large corrections could outweigh the benefits of accessing the desired quantities at zero momentum. 
To this end, 
we have extended the method to the computation of moments of form factors, 
such as charge radii, 
and investigated the theoretical framework necessary to deduce the finite volume corrections%
~\cite{Tiburzi:2014yra}.
We review these developments here.

\section{Zero-Momentum Method for Radii}

To extend the Rome method to moments of form factors, 
we focus on a particularly simple example, 
namely that of the pion form factor. 
The pion form factor is completely connected at the quark level%
\footnote{
Our method to determine the pion charge radius straightforwardly generalizes to the quark-connected part of the nucleon charge radius. 
} 
owing to a combination of charge-conjugation and isospin invariance%
~\cite{Draper:1988bp,Bunton:2006va}. 
The pion charge radius, 
moreover, 
is a well-tested lattice calculation, 
and so becomes an ideal proving ground for techniques at zero momentum.  
In this section, 
we begin by working at infinite volume.

The charge radius appears in the momentum-transfer expansion of the pion form factor. 
The form factor parameterizes the current matrix element in the pion
\begin{equation}
\langle \pi^+ (\vec{p} \,') | J_\mu | \pi^+ (\vec{p}) \rangle
= 
e
(p' + p)_\mu
F(q^2)
,\end{equation}
with 
$q_\mu = (p' - p)_\mu$
is the momentum transfer. 
The form factor has the momentum expansion 
$F(q^2) = 1 - \frac{1}{6} < r^2> q^2 + \cdots$, 
with the first non-trivial coefficient,
$< r^2>$, 
defined to be the charge radius. 
Notice that throughout, 
we work in Euclidean space. 
The current matrix element can be extracted from the time component of the lattice three-point function, 
which we generically write as 
$C_4 ( \vec{p} \, ' , \vec{p} \, | x_4, y_4 )$, 
where 
$x_4$ is the source-sink separation, 
and 
$y_4$ is the current insertion time. 
In the rest frame, 
we have 
$\vec{p} = \vec{0}$, 
and
$\vec{p} \, ' = \vec{q}$. 
With this choice of frame, 
the four-momentum transfer is 
$q^2 = \vec{q} {}^2 [ 1 + \mathcal{O} ( \vec{q} {}^2 / m_\pi^2 ) ]$. 
In the limit of long Euclidean time separation between source and sink, 
and between current insertion time and source, 
we have the expected behavior of the pion current correlation function
\begin{equation}
C_4 (\vec{q}, \vec{0} \,| x_4, y_4)
=
i (E' + m_\pi) F(q^2) |Z|^2 \frac{e^{ - E' (x_4 - y_4)} e^{ - m_\pi y_4}}{2 E' \, 2 m_\pi}
,\end{equation}
where the energy of the final state pion is 
$E' = \sqrt{\vec{q} \, {}^2 + m_\pi^2}$.

%
%
\begin{figure}
\begin{center}
\includegraphics[width=0.4\textwidth]{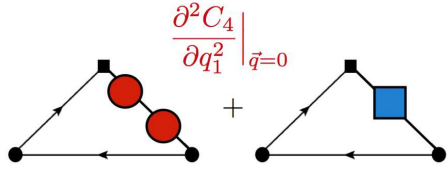}
\caption{
Graphical depiction of the quark-level correlation function for the Taylor series coefficient in the rest frame. 
The two red circles and blue square represent the second-order terms in the Taylor series expansion of the quark propagator about zero momentum.
This correlation function is sensitive to the charge radius, but there are additional contributions. 
}
\label{f:rest}
\end{center}
\end{figure}
%
%

Momentum derivatives of this correlation function can be used to try and isolate the charge radius, 
however, 
there are additional terms produced upon differentiation. 
The second Taylor series coefficient has the form
\begin{equation}
- \frac{3}{C_4}
\frac{\partial^2 C_4}{\partial q_1^2}
\Bigg|_{\vec{q} = \vec{0}}
=
< r^2 > - \frac{3}{2} m_\pi^{ -2} + ( x_4 - y_4) m_\pi^{-1}
.\end{equation}
Computation of the Taylor series coefficient can be carried out as depicted in Fig.~\ref{f:rest}, 
and requires replacing quark propagators by second-order terms in their derivative expansion about zero momentum, 
\begin{equation}
S(\vec{k} \,) 
= 
S(0)
- 
S(0) 
\vec{k} \cdot \vec{V} 
S(0) 
+ 
S(0)
\vec{k} \cdot \vec{V} \,
S(0)
\vec{k} \cdot \vec{V} \,
S(0)
- 
\frac{1}{2}
S(0) \vec{k} \cdot  \overset{\leftrightarrow}{T} \cdot \vec{k} \, S(0)
+ 
\cdots,
\label{eq:propexp}
\end{equation}
where
$V_\mu$
is the point-split vector current, 
and 
$T_{\mu \nu}$
is a tadpole current. 
While momentum derivatives of the rest-frame current correlation function produces the charge radius, 
there is a problematic additive contribution that depends on the Euclidean time separation. This additive term will dominate over the charge radius, 
but could be removed by carefully studying the current insertion-to-sink time dependence of the correlation function. 
There is an additional worry for derivatives of the rest-frame current correlation function. 
At finite lattice spacing, 
there is the possibility of divergent terms arising from two identical vector currents at the same space-time point, 
see%
~\cite{Aubin:2013daa,Horch:2013lla}. 
These drawbacks obstruct extraction of the charge radius from the Taylor series expansion of the rest-frame current matrix element.

%
%
\begin{figure}
\begin{center}
\includegraphics[width=0.22\textwidth]{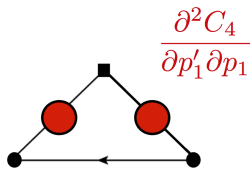}
\caption{
Depiction of the correlation function for the Taylor series coefficient in an arbitrary frame.
This correlator is directly proportional to the charge radius.  
}
\label{f:arbitrary}
\end{center}
\end{figure}
%
%

To remedy the situation, 
we turn our attention to the current matrix element in an arbitrary frame. 
Here, 
one has two independent momenta to vary. 
In an arbitrary frame, 
the temporal component of the current matrix element has the expected behavior
\begin{equation}
C_4 (\vec{p} \, ', \vec{p} \,| x_4, y_4)
=
i (E' + E) F(q^2) |Z|^2 \frac{e^{ - E' (x_4 - y_4)} e^{ - E y_4}}{2 E' \, 2 E}
,\end{equation}
provided the Euclidean time separations are large enough to ensure ground-state saturation. 
With the momentum transfer now given by
$q^2 = 2 [ E' E - m_\pi^2 - \vec{p} \, ' \cdot \vec{p} \, ]$, 
we can cleanly isolate the charge radius by computing 
\begin{equation}
- \frac{3}{C_4}
\frac{\partial^2 C_4}{\partial p'_1 \partial p_1}
\Bigg|_{\vec{p} \,' = \vec{p} = \vec{0}}
=
< r^2 > 
.\end{equation}
At the quark level, 
the corresponding Taylor series correlation function is shown in Fig.~\ref{f:arbitrary}. 
Only the point-split vector current insertion is needed, 
and there can be no divergent contact terms at finite lattice spacing due to the temporal separation of the initial- and final-state quark propagators. 
We also find that two higher moments of the charge distribution, 
namely
$<r^4>$
and 
$<r^6>$,
can be cleanly accessed by taking further momentum derivatives with respect to the remaining spatial components of the momenta. 
Having described a method, 
albeit computationally expensive, 
to determine the charge radius directly, 
a natural question to ask is whether potentially large finite volume corrections spoil the method from the outset. 
To answer this question for moments of the charge distribution of the pion, 
we turn to chiral perturbation theory.

\section{Finite Volume Effects}

To determine the effect of a finite volume on the Rome method, 
we must develop a framework to derive the Taylor series coefficients on a fixed-size lattice. 
On a lattice of fixed size, 
quarks subjected to periodic boundary conditions have quantized momenta, 
and consequently differentiation with respect to integer mode numbers will not yield a Taylor series expansion. 
The way to derive the Taylor series expansion of correlation functions on a lattice of fixed size 
$L$ 
is to introduce active and spectator quarks. 
The active quarks are essentially the momentum carriers, 
and they can carry continuous momenta if they are subjected to partially twisted boundary conditions. 
For a boundary condition of the form
$\psi (x+ L ) = e^{ i \theta} \psi(x)$, 
the corresponding momentum is
$p = \theta / L$. 
To derive the Taylor series coefficients in a finite volume, 
one uses the equivalence
$p \frac{\partial}{\partial p} 
= 
\theta \frac{\partial}{\partial \theta}$. 
Notice that the Taylor coefficients are evaluated a zero momentum, 
i.e.~vanishing twist angle. 
As a result, 
one still computes these coefficients in the way prescribed by the Rome method. 
To address finite volume effects, 
however, 
we must understand the underlying theoretical framework. 
The framework justifying a Taylor series expansion in momentum on a fixed-size lattice is that of partially twisted boundary conditions on the active quarks.

%
%
\begin{figure}
\begin{center}
\includegraphics[width=0.3\textwidth]{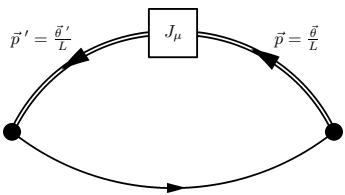}
\caption{
Quark-level contractions required to determine the pion current matrix element in an arbitrary frame. 
The differing twisted boundary conditions allow one to perform the Taylor series expansion in momentum on a fixed-size lattice, but force one to distinguish between two active quarks, 
and a spectator quark.  
}
\label{f:foggy}
\end{center}
\end{figure}
%
%

To deduce the finite volume effect on the  method developed here to extract the pion charge radius, 
we must enlarge the flavor group from 
$SU(2)$ 
to 
$SU(5|3)$. 
Formal discussion of the construction and foundations of partially quenched chiral perturbation theory can be found in%
~\cite{Sharpe:2000bc,Sharpe:2001fh}.
The reason for the extension to a graded Lie group is as follows. 
In addition to the spectator quark 
(which can be chosen to be the down quark), 
there must be two distinguishable active quarks: 
an initial-state up quark carrying momentum 
$\vec{p} = \vec{\theta} / L$, 
and a final-state up quark carrying momentum 
$\vec{p} \, ' = \vec{\theta} \, ' / L$. 
These are all valence quarks; 
and, 
to make this identification, 
we  must introduce corresponding ghost quarks. 
The two sea quarks of the 
$SU(5|3)$
flavor group remain periodic. 
The pion form factor in an arbitrary frame can be computed at finite volume from the flavor-changing 
current shown in 
Fig.~\ref{f:foggy}. 
Here the flavor transition is between the two active quarks, 
and so they only differ by their differing boundary conditions. 
We utilize the 
so-called 
$p$-regime of chiral perturbation theory in our investigation, 
where the zero modes of the pion field remain weakly coupled%
~\cite{Gasser:1987zq}. 
One might worry that the partially twisted scenario might introduce enhanced finite volume effects due to unitarity violations. 
Such enhancement occurs from the double poles present in hairpin propagators. 
These hairpins arise from propagation of flavor-neutral Goldstone modes, 
and it is important for us to note that flavor-neutral states are not altered by twisted boundary conditions, 
the effect from the quark is exactly cancelled by the antiquark in the flavor-neutral meson. 
As a result, 
flavor neutral propagators have exactly the same form as in QCD, 
provided the valence and sea quark masses are made degenerate.

To compute the finite volume effect on the extraction of moments of the pion charge distribution at zero momentum, 
we take the requisite derivatives of the partially twisted pion current matrix element computed in an arbitrary frame. 
Notice that at finite volume, 
there is no frame independence of current matrix element. 
As a consequence, 
existing partially twisted pion current matrix elements computed at finite volume in the rest frame%
~\cite{Jiang:2006gna},
and Briet frame%
~\cite{Jiang:2008te}
cannot be utilized. 
The computation in an arbitrary frame, 
however, 
properly reduces to these two previously investigated cases.

\begin{figure}
\begin{center}
\includegraphics[width=0.475\textwidth]{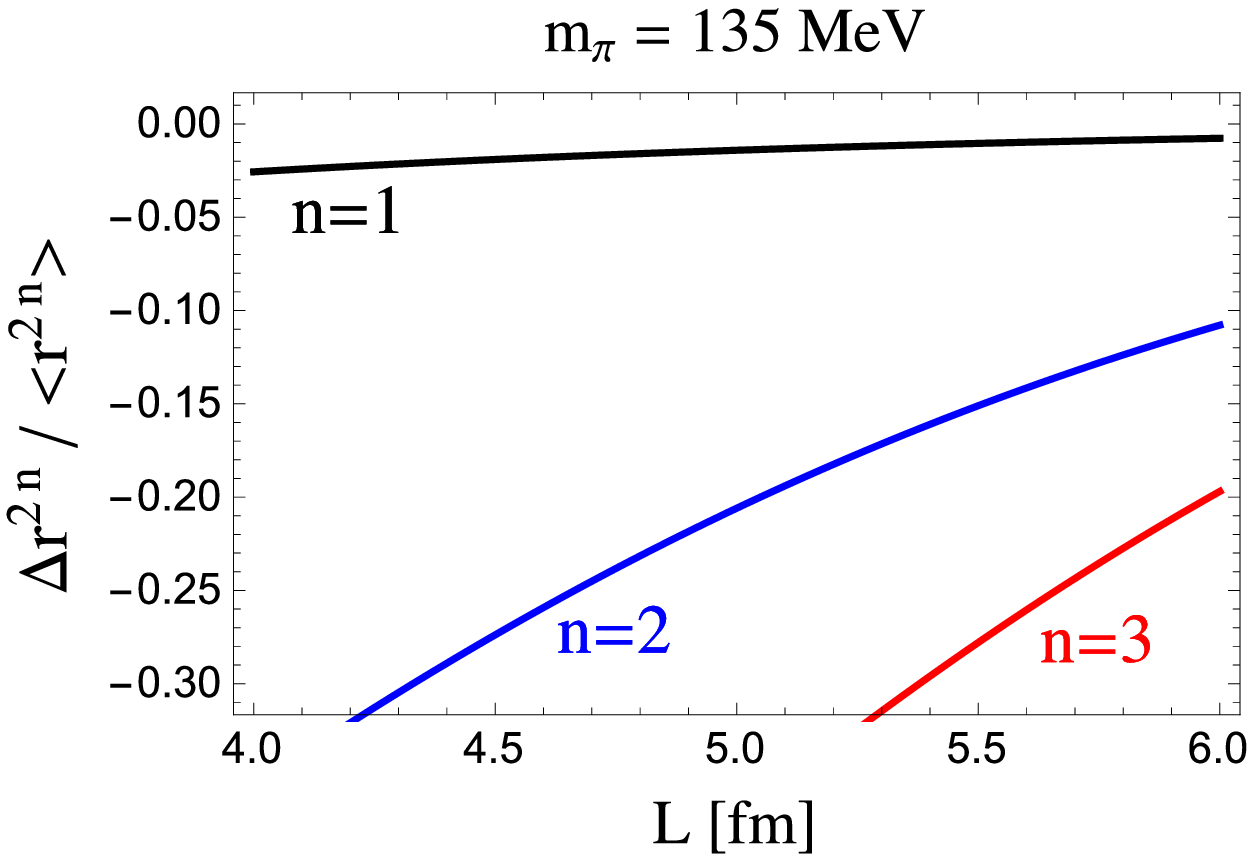}
\includegraphics[width=0.475\textwidth]{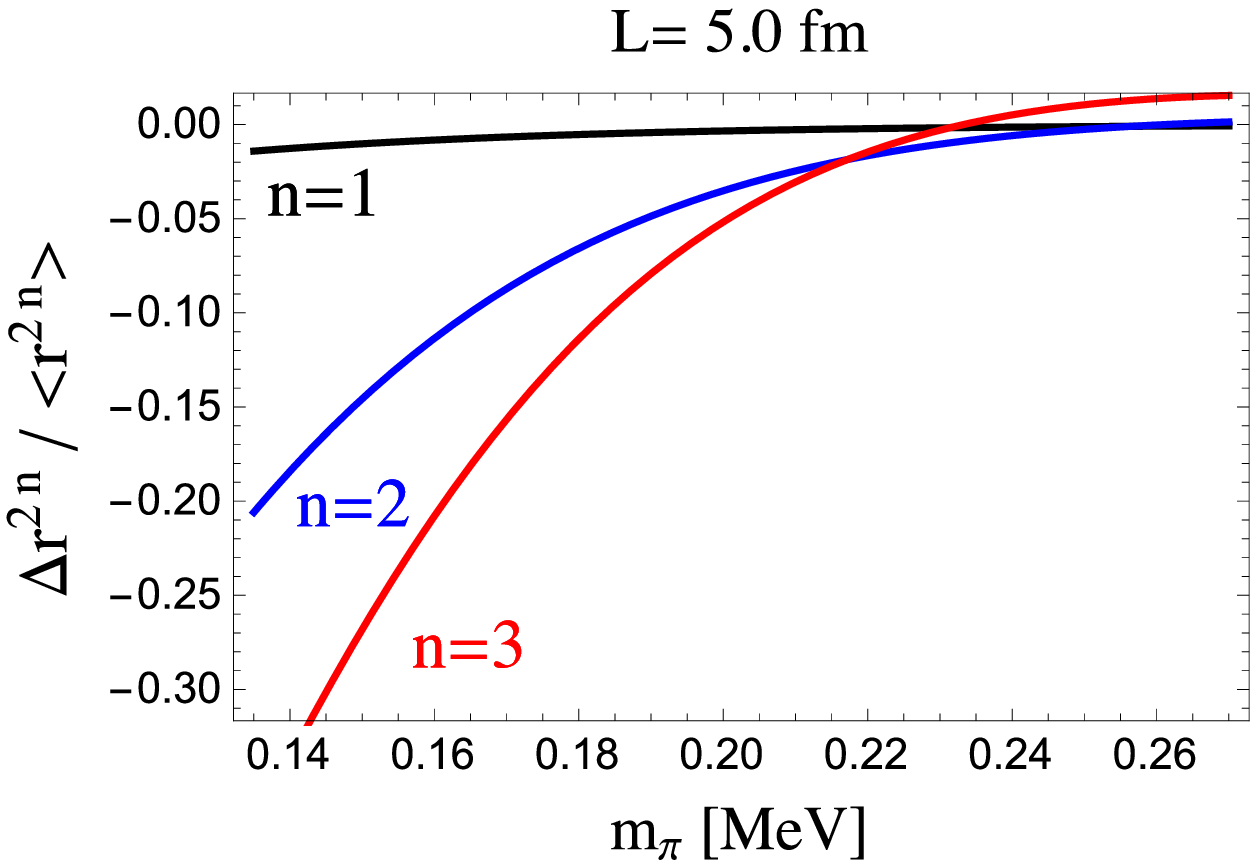}
\caption{
Comparison of finite volume effects on the Rome method as deduced from partially twisted chiral perturbation theory. 
Shown in the left panel is the relative difference of the first three moments of the pion charge distribution as a function of the lattice size 
$L$ 
at the physical pion mass. 
In the right panel, 
by contrast, 
we fix the lattice size to 
$L = 5.0 \, \texttt{fm}$,
and plot the relative differences as a function of the pion mass. 
}
\label{f:FV}
\end{center}
\end{figure}

The finite volume computation can be summarized in the following expression. 
The matrix element of the temporal component of the current takes the form
\begin{equation}
\Delta \mathcal{M}_4 
\equiv 
\mathcal{M}_4 (L) - \mathcal{M}_4 (\infty)
=
(p'+p)_4 \, 
\Delta F 
+ 
q_4 \, \Delta G
.\end{equation}
At finite volume with twisted boundary conditions, 
there is essentially an extra form factor, 
$\Delta G$. 
Such an additional contribution was originally found in%
~\cite{Jiang:2006gna}, 
while the origin of this term has now been linked to the necessity of maintaining the Ward-Takahashi identity%
~\cite{Bijnens:2014yya}. 
In an arbitrary frame, 
the functional form of 
$\Delta G$
is quite complicated. 
Fortunately contributions from 
$\Delta G$
drop out of all Taylor series coefficients determined in an arbitrary frame. 
As a result, 
we can deduce the finite volume effect on moments of the pion charge distribution by taking the required
twist-angle derivatives of 
$\Delta F$. 
The size of such finite volume corrections is assessed in Fig.~\ref{f:FV}, 
where the relative difference in finite and infinite volume contributions is plotted as a function of the lattice 
size 
$L$ 
and pion mass
$m_\pi$. 
For the infinite volume values, 
we use the experimental radius and predictions from one-loop chiral perturbation theory%
~\cite{Gasser:1984ux}.
While the higher moments are power-law enhanced in a finite volume, 
\begin{equation}
\Delta r^{2 n} \sim L^{2( n -1)} \sqrt{m_\pi L} \, \, e^{ - m_\pi L}
,\end{equation}
the overall scaling remains exponential. 
For the charge radius in particular, 
the zero momentum method is expected to introduce finite volume corrections that are less than a few percent.

%
%
%
%

\section{Summary}

Above we review the novel Rome method, 
which has been proposed to overcome large extrapolations to vanishing momentum. 
We discuss extension of the method to the case of radii, 
focusing on the pion charge radius for simplicity. 
The extension to radii is potentially problematic: 
momentum derivatives of rest-frame current matrix elements introduce power-law Euclidean time dependence that  complicates extraction of the radius. 
Divergent contact terms, 
moreover, 
are expected to be encountered in this approach. 
By contrast, 
momentum differentiation of the current matrix element in an arbitrary frame leads to different Taylor-coefficient correlation functions. 
These are free from power-law Euclidean time contamination, and free from divergent contact terms. 
Because the required correlation functions appear to require extensive computational resources, 
we investigate the impact of finite volume corrections in advance of costly numerical explorations. 
By modifying chiral perturbation theory, 
we are able to compute finite volume corrections to the Rome method.
In the process, 
we ascertain that finite volume corrections to the pion charge radius are generally at the percent level on current-size lattices, 
even at the physical pion mass. 
A straightforward generalization of the method is required to investigate moments of the electric and magnetic form factors of the nucleon, 
with magnetic quantities anticiapted to be more sensitive to the volume. 
The method, 
however, 
is only practicable for the connected part of nucleon properties. 
For disconnected current insertions, 
gluons are required as momentum carriers,
and we have been unable to extend the Rome method to this case. 
It remains to be seen whether an alternative approach exists.

\end{document}